\documentclass[reprint,superscriptaddress,amsmath,amssymb,aps,pra]{revtex4-1}

\usepackage{graphicx,epstopdf}
\usepackage{dcolumn}
\usepackage{bm}
\usepackage{hyperref}

\usepackage{multirow}
\usepackage{threeparttable}
\usepackage{color}
\usepackage{verbatim}
\usepackage{subfigure}
\usepackage{setspace}

\usepackage[svgnames]{xcolor}
\usepackage[normalem]{ulem}

\begin{document}
\title{Detecting Entanglement in Unfaithful States}
\author{Yongtao Zhan}
\email{yongtao.zhan@mail.utoronto.ca}
\affiliation{Department of Electrical and Computer Engineering, University of Toronto, Toronto, Ontario, M5S 3G4, Canada}

\author{Hoi-Kwong Lo}

\affiliation{Department of Electrical and Computer Engineering, University of Toronto, Toronto, Ontario, M5S 3G4, Canada}
\affiliation{Department of Physics, University of Hong Kong, Pokfulam, Hong Kong}

\begin{abstract}
Entanglement witness is an effective method to detect entanglement in unknown states without doing full tomography. One of the most widespread schemes for witnessing entanglement is measuring its fidelity with respect to a pure entangled state. Recently, a large class of states whose entanglement cannot be detected with the fidelity witness has been discovered in Phys.Rev.Lett. \textbf{124},200502(2020). They are called unfaithful states. In this paper we propose a new way to detect entanglement by calculating the lower bound of entanglement using measurement results. Numerical simulation shows that our method can detect entanglement in unfaithful states with a small number of measurements. Moreover, we generalize our scheme to multipartite states and show that it can tolerate higher noise than previous entanglement witness operators with same number of measurement settings.

\end{abstract}

\maketitle


The concept of entanglement has played a crucial role in the development of quantum physics. The Einstein, Podolsky and Rosen effect~\cite{PhysRev.47.777} has received world-wide attention and has been a hot topic of research for decades. Entanglement has come to be recognized
as a novel resource that may be used to perform
tasks that are either impossible or very inefficient in the classical realm, such as Quantum Key Distribution (QKD)~\cite{RevModPhys.92.025002}, quantum superdense coding~\cite{PhysRevLett.69.2881}, quantum teleportation~\cite{PhysRevLett.70.1895}, and quantum computing~\cite{PhysRevA.68.022312}. These developments have provided the
seed for the development of modern quantum information science.

It is important for us to know whether a state prepared in an experiment is entangled or not. A trivial way -- performing full tomography of the state -- requires a large amount of measurements. With an a $N$-qubit state, for example, we would need $2^{2N}$ measurements to determine all the matrix elements of the density matrix $\rho$. This means that the number of measurements grows exponentially with the particle number $N$.

\emph{Entanglement witness} is an efficient scheme to detect entanglement in an unknown state, requiring a much smaller number of measurements compared to full tomography~\cite{natrevphy}. The set of separable states is a convex subset of the Hilbert space. The Hahn–Banach theorem guarantees that there exists a hyperplane for every entangled state $\rho$ that separates this state from the separable set. These hyperplanes correspond to observables $\mathcal{W}$, such that $Tr(\mathcal{W}\rho) < 0$, whereas $Tr(\mathcal{W}\sigma) \geq 0$ for all separable states $\sigma$.  If the measured value $\langle \mathcal{W}\rangle_{\rho}$ is positive, we cannot tell whether the state is entangled or not. However, we can know for sure that $\rho_{exp}$ is an entangled state if the measurement produces a negative expectation value. Such an  observable $\mathcal{W}$ is called an entanglement witness operator. Although $\mathcal{W}$ exists for each entangled state $\rho$, it is formidable to construct a useful witness operator without specific information about the state produced in the experiment.

However, when the underlying state is expected to be close to a target entangled state $|\psi\rangle$, entanglement can be witnessed by measuring the fidelity between the two states. This method is called the fidelity witness. A fidelity witness operator that detects entanglement of a pure state $|\psi\rangle$ (and of states that are close to $|\psi\rangle$) is given by \cite{PhysRevLett.92.087902}
\begin{equation}\mathcal{W}=\alpha \mathbb{I}-|\psi\rangle\langle\psi|\label{11}.\end{equation} 
The factor $\alpha$ must be chosen such that all
separable systems lead to non-negative expectation values
(a method for determining $\alpha$ is presented in Ref.\cite{PhysRevLett.92.087902}).
Fidelity witnesses have been widely studied both theoretically and experimentally~\cite{PhysRevLett.92.087902,PhysRevLett.94.060501,PhysRevA.72.022340,PhysRevLett.100.210501,PhysRevLett.103.020504,PhysRevLett.95.210502,nphys,PhysRevLett.103.160401,natureexp}.

Unfortunately, a recent paper shows that the fidelity witness is severely limited in its ability in detecting entanglement and, in fact, the entanglement in a large class of states, namely the \emph{unfaithful states}, cannot be detected by the fidelity witness\cite{PhysRevLett.124.200502}. For high-dimensional systems ($d>2$), they find, surprisingly, that almost no bipartite entangled state can be detected using the fidelity witness. Their work is a warning against the blind application of fidelity-type entanglement witnesses and shows the importance of detecting entanglement in unfaithful states.

In this Letter we present a generalized entanglement witness which can detect entanglement in unfaithful states.  By estimating the lower bound of certain entanglement measure $E(\rho)$(e.g. geometric measure) based on a set of measured values of observables using the method introduced in Ref.\cite{PhysRevLett.98.110502}, we know for sure that the state is entangled if we get a positive lower bound, given the entanglement measure $E(\rho)=0$ if $\rho$ is a separable state. Our method can detect entanglement in unfaithful states, which surpass the capability of fidelity witness.

Our method can be extended to detect genuine entanglement in multipartite states by estimating the lower bound of generalized geometric measure(GGM). Similar to the bipartite case, our method can detect entanglement beyond fidelity witness. As fidelity witness usually requires a large number of measurements in multipartite cases, fidelity-type witnesses, which use fewer measurements, are usually used in experiment. We also show that the capability of our method surpass fidelity-type witnesses given the same measurements.

Unlike previous entanglement witness operators, there is no restrictions on the type and number of measurements in our method. The type of measurements can be customized to adapt various experimental restrictions. The number of measurements can be added or reduced to reach different capability of detecting entanglement.

Our scheme is summarized as follows: We measure a set of Hermitian operators $A_1,A_2,\dots,A_n$ in an experiment and use the estimated expectation values $a_1,a_2,\dots,a_n$ to calculate the lower bound of entanglement of the state $\rho$
\begin{equation}
\varepsilon\left(a_{1}, \ldots, a_{n}\right)=\inf _{\rho}\left\{E(\rho) \mid \operatorname{Tr}\left(\rho {A}_{k}\right)=a_{k}\right\},
\end{equation}
where $E(\rho)$ is a certain entanglement measure and the infimum is understood as the infimum over all states compatible with the data $a_k=\operatorname{Tr}(A_k\rho)$. In this paper, we choose $E(\rho)$ to be the geometric measure of entanglement $E_G$ for bipartite states\cite{PhysRevA.68.042307}. The geometric measure $E_G$ is positive only when a bipartite state is entangled. If we get a positive value of the lower bound of entanglement $\varepsilon\left(a_{1}, \ldots, a_{n}\right)$, we know for sure that the state is entangled. On the other hand, if the lower bound is zero, we cannot tell whether a state is entangled or not.

The method for estimating the lower bound of entanglement is introduced in Ref.\cite{PhysRevLett.98.110502}. Here we briefly summarize their idea. The detailed estimation process and the algorithms are given in the Appendix. 

\begin{figure}
    \centering
    \includegraphics[width=0.47\textwidth]{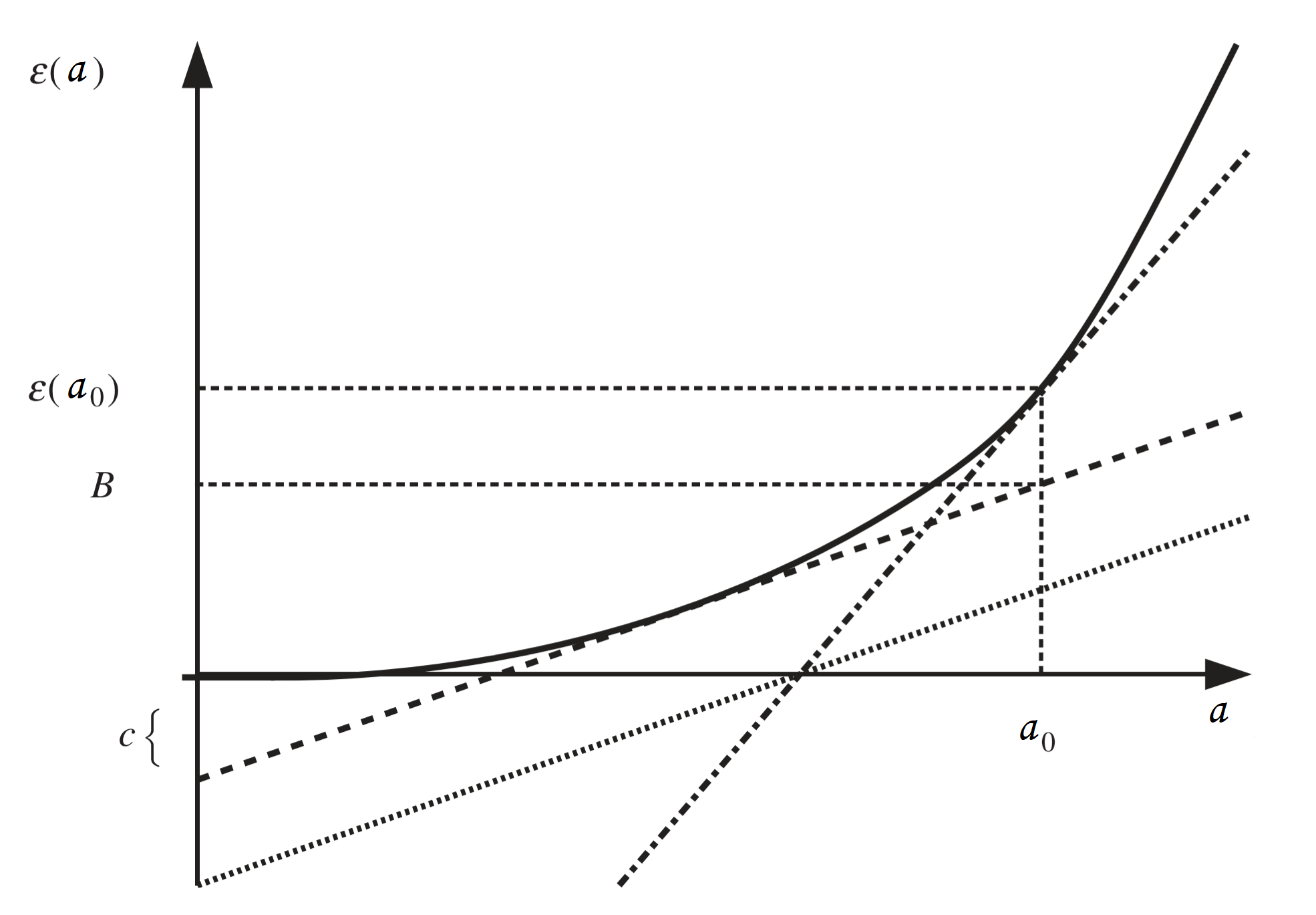}
    \caption{A schematic view of the method of estimating entanglement lower bound. This figure is reproduced from Ref.\cite{PhysRevLett.98.110502}. The lower bound $\varepsilon(w)$ is a convex function. We assume that $w_0$ is the measured expectation value. The dotted line corresponds to a general estimation. First we perform the minimization on the intercept $c$ and the linear bound becomes the dashed line. We can get a better estimation of the entanglement lower bound, which is $B$. By varying the slope $r$ we arrive at the dash-dotted line, which gives the best estimation of the lower bound $\varepsilon(w_0)$. }
    \label{mfig1}
\end{figure}

We consider linear bounds of the type
\begin{equation}\varepsilon(a) \geq  r \cdot a-c, \label{m66}\end{equation}
where $r$ and $a$ are both $n$-dimensional vectors and the dot denotes the dot product. For the one-dimensional case, the two bounds are plotted in Figure~\ref{mfig1}. To find the best estimation of the entanglement lower bound $\varepsilon(a)$, the intercept $c$ should be as small as possible. By the definition of $\varepsilon(a)$, we note that Eq.~(\ref{m66}) can be expressed as $E(\rho)\geq r \cdot a-c$, so we arrive at the inequality for $c$.
\begin{equation}c \geq \sum_{k} r_{k} \operatorname{Tr}\left(\rho A_{k}\right)-E(\rho)\end{equation}
This indicates that we should perform maximization over the whole state space and choose $c$ to be the supremum of the right-hand side (RHS), which only depends on the operator $\mathcal{A}=\sum_k r_k A_k$. Thus the supremum of the RHS can be expressed as the function $\hat{E}(\mathcal{A})$, where
\begin{equation}\hat{E}(\mathcal{A})=\sup _{\rho}\{\operatorname{Tr}(\rho \mathcal{A})-E(\rho)\}\label{1515}.\end{equation}
After obtaining the minimum value of $c$, we vary the slope $r$ to arrive at the best possible bound $\varepsilon(a)$, as we have shown in Fig.~\ref{mfig1}:
\begin{equation}\varepsilon(a)=\sup _{r}\left\{\sum_k r_k a_k-\hat{E}\left(\sum_{k} r_{k} A_{k}\right)\right\}\label{m99}\end{equation}
Note that Eq.~(\ref{m99}) shows that $\varepsilon(a)$ is the Legendre transformation of the function $\hat{\varepsilon}(r)=\hat{E}(\sum_k r_k A_k)$.

\begin{figure}
    \centering
    \includegraphics[width=0.5\textwidth]{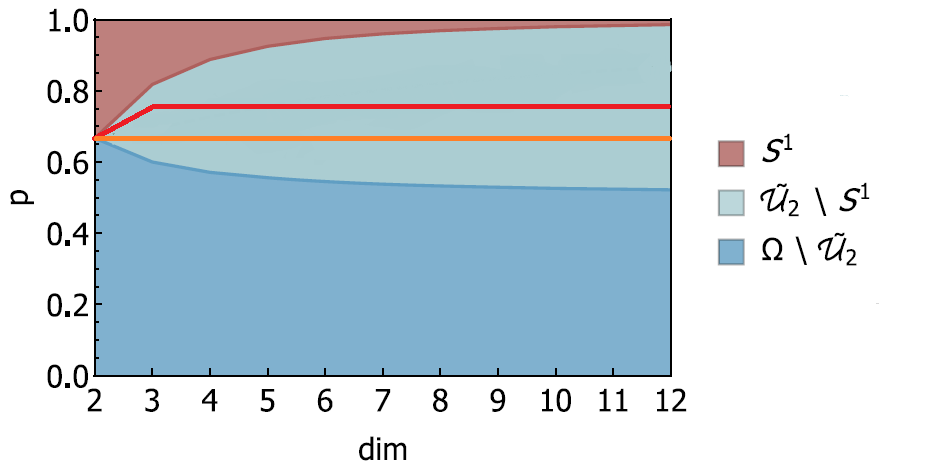}
    \caption{The noise tolerance of fidelity witnesses and our method. The red region corresponds to the set of separable states. The green region indicates the set of unfaithful and entangled states. The blue region indicates the set of faithful states, whose entanglement can be detected using fidelity witness. The symbols $\mathcal{S}^1$, $\tilde{\mathcal{U}}_2\backslash\mathcal{S}^1$,$\Omega\backslash\tilde{\mathcal{U}}_2$ denotes the approximation of the three set of states using SDP.(See Appendix or Ref.\cite{PhysRevLett.124.200502} for how the approximation was done) The blue line is the bound of noise tolerance of any fidelity witness. The orange line is the noise tolerance of our method using only three measurements, $X \otimes X$,$ Y\otimes Y$,$ Z\otimes Z$. If we add one more measurement -- $M\otimes M$ -- the noise tolerance is higher, which is depicted by the orange line. Our scheme achieves a big improvement over the fidelity witness.}
    \label{fig2}
\end{figure}

Our new scheme can be used to detect entanglement in bipartite unfaithful state. For a convenient comparison, we take the Bell state $|\Psi_2\rangle=(|00\rangle+|11\rangle)/\sqrt{2}$ embedded in a $d \times d$-dimensional Hilbert space (Eq.(\ref{33})) as an example 
\begin{equation}
    \rho=p\frac{\mathbb{I}}{d^2}+(1-p)|\Psi_2\rangle\langle\Psi_2|
    \label{33}
\end{equation}
where $p$ is the amount of white noise. It is easy to prepare such a state by drawing a random state from either a completely mixed state or the state $| \Psi_2 \rangle$ with a probability $p$ and $1-p$, respectively. In Ref.\cite{PhysRevLett.122.120501}, the authors analyze the unfaithfulness of this mixed state. They find there exists a large parameter regime where the state $\rho$ is entangled and unfaithful for $d>2$. This phenomenon is shown in Fig.\ref{fig2}. 

For a convenient comparison, we use the same measurements with the fidelity witness to estimate the lower bound of entanglement of the state, which are $X\otimes X, Y\otimes Y,$ and $Z\otimes Z$. The operators $X$, $Y$ and $Z$ are defined as follows:

\[
X=\left[
\begin{array}{cc|c} 
   0  &    1     &         \\
       1     & 0    &     \\
    \hline              
             &          &  0 
\end{array}
\right]
,
Y=\left[
\begin{array}{cc|c} 
   0  &    -i     &         \\
       i     & 0    &     \\
    \hline              
             &          &  0 
\end{array}
\right]
,
Z=\left[
\begin{array}{cc|c} 
   1  &    0    &         \\
       0     & -1    &     \\
    \hline              
             &          &  0 
\end{array}
\right]
\]
They are the Pauli operators in the subspace spanned by $|0\rangle$ and $|1\rangle$. These operators are easy to measure in an experiment. For example, for qudits encoded in photon paths, we just have to measure the first and second path to obtain the outcomes $X\otimes X, Y\otimes Y, Z\otimes Z$. It is easy to check that $X\otimes X$,$Y\otimes Y$ and $Z\otimes Z$ are stabilizers of $|\Psi_2\rangle$. We use expectation values of the three stabilizers to calculate the lower bound of geometric entanglement. For the the state  $\rho=p{\mathbb{I}}/{d^2}+(1-p)|\Psi_2\rangle\langle\Psi_2|$, the expectation values are $\langle X \otimes X \rangle=\langle Y \otimes Y \rangle=\langle Z \otimes Z \rangle=1-p$. For any dimension $d$, the entanglement lower bound is positive when $p<2/3$, at which point we know that the state is entangled. For $p>2/3$, the lower bound is zero and we cannot tell if the state is entangled. Therefore, the white noise tolerance threshold of our method is $p=2/3$. 

We can add more measurements to obtain a better estimation of the lower bound of entanglement. For example, a measurement of $M\otimes M$ can be added when $d\geq3$, where the operator $M$ reads
\[
M=\left[
\begin{array}{ccc|c} 
   1  &    0    & 0 &        \\
       0     & 0 &  0  &     \\
       0 & 0 & -1 \\
    \hline              
             &    &      &  0 
\end{array}
\right]
\]
This is a Pauli-$Z$ operator in the subspace spanned by $|0\rangle$ and $|2\rangle$. Hence we use four expectation values -- $\langle X \otimes X \rangle$,$\langle Y \otimes Y \rangle$, $\langle Z \otimes Z \rangle$ and $\langle M \otimes M \rangle$ -- to estimate the lower bound of entanglement. For any dimension $d>2$,  the lower bound is positive when $p<3/4$. This means the noise tolerance of our method is $p=3/4$ with four measurements.

The numerical results are shown in Fig.\ref{fig2}. The blue line is the noise tolerance bound of any fidelity witness. The orange line is the noise tolerance bound when three measurements $X \otimes X,Y\otimes Y, Z\otimes Z$ are used to detect entanglement.  When we add one more measurement, $M\otimes M$ for $d\geq3$, the noise tolerance bound becomes higher, which is shown by the red line. If we use more measurements, the noise tolerance can be higher. Therefore our method can be useful to detect entanglement in unfaithful bipartite states.

Furthermore, our scheme can be extended to the case of multipartite stabilizer states (e.g. cluster states) and non-stabilizer states (e.g. $W$ state). In order to detect genuine entanglement in multipartite states, We choose the entanglement measure $E(\rho)$  to be the generalized geometric measure (GGM) \cite{PhysRevA.94.022336,PhysRevA.81.012308}. The GGM of a pure state $|\psi\rangle$ is defined by
\begin{equation}
E_{GGM}(|\psi\rangle)=1-\sup _{|\phi\rangle}  |\langle\phi \mid \psi\rangle|^2
\end{equation}
where the optimization is done over all biseparable states $|\phi\rangle$. The GGM of mixed states is defined using the convex roof construction\cite{PhysRevA.94.022336,PhysRevA.81.012308}. The GGM is positive only when the state is genuine entangled. So the state $\rho$ must carry genuine multipartite entanglement if the lower bound of GGM of state is positive. Moreover, we find that our scheme can tolerate higher noise than the previously constructed witness operators~\cite{PhysRevA.72.022340,npj,PhysRevLett.94.060501,Guhne_2009} while using the same number of measurement settings.

Here we show our scheme can be used to detect genuine entanglement in cluster states.  Although fidelity witnesses can be used to detect genuine entanglement, they are not widely used in multipartite states because measuring the fidelity requires a large number of single-qubit measurements. In order to solve this problem, a large class of witness operators called \emph{fidelity-type witness} has been derived~\cite{PhysRevA.72.022340,Guhne_2009}. (See Appendix for how fidelity-type witness is derived.)

A \emph{fidelity-type witness} operator is presented in \cite{PhysRevLett.94.060501,PhysRevA.72.022340} to detect entanglement in linear qubit cluster states:
\begin{equation}
\mathcal{W}_c:=3 \mathbb{I}-2\left[\prod_{\text {even } k} \frac{S_{k}+\mathbb{I}}{2}+\prod_{\text {odd } k} \frac{S_{k}+\mathbb{I}}{2}\right] \label{2121}
\end{equation}
where $S_k$ refers to the $k^\text{th}$ stabilizer generator. These witness operators are robust against noise and require only two local measurement settings when used in an experiment, independent of the number of qubits, which is a big improvement over the fidelity witness operator in Eq. (\ref{11}).
\begin{figure}
    \centering
    \includegraphics[width=0.5\textwidth]{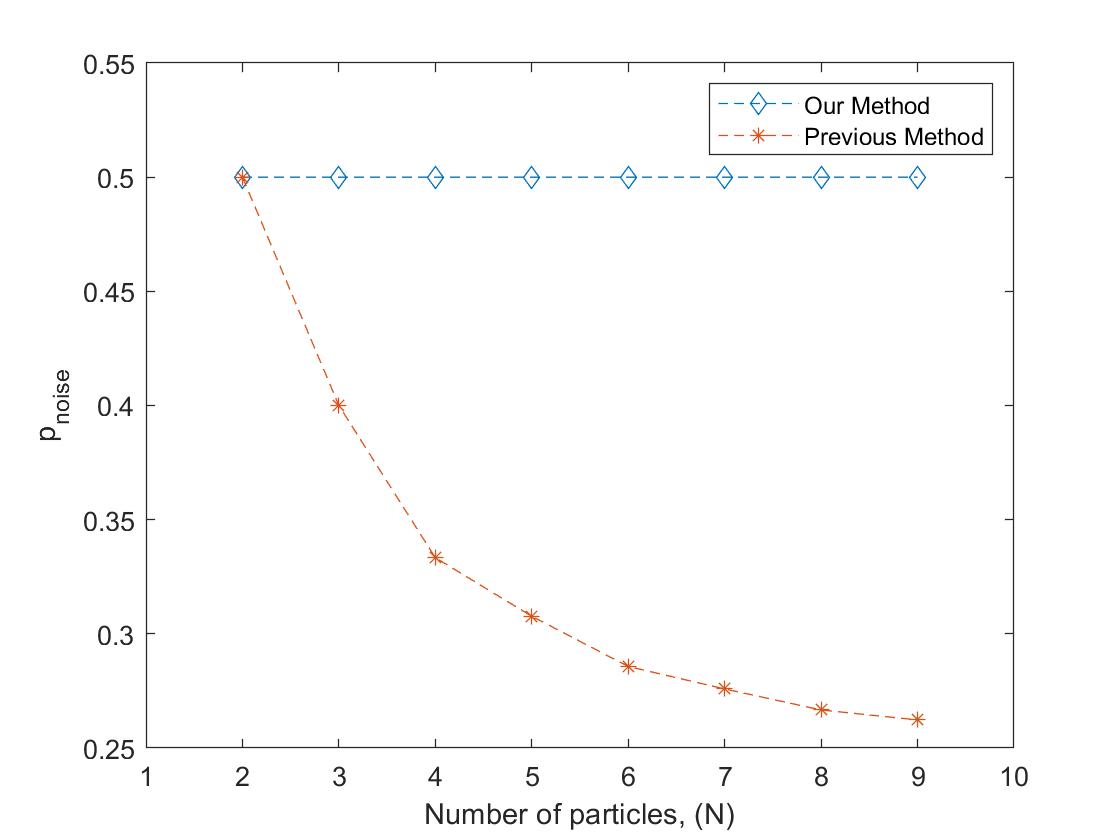}
    \caption{The noise tolerance of our method and the witness operator $\mathcal{W}_1$ as a function of $N$, the number of qubits in the cluster state. The noise tolerance of our method is about 0.5 for any number of particles, while the noise tolerance of $\mathcal{W}_1$, which is the best witness operator that has ever been derived, is close to 0.25 for large number $N$. }
    \label{fig3}
\end{figure}
When the cluster state $|\phi_{N,2}\rangle$ is mixed with certain amount of white noise $p_\text{noise}$, the state becomes
\begin{equation}\rho=p_{\text{noise}} \mathbb{I}/2^N +(1-p_{\text{noise}})|\phi_{N,2}\rangle\langle\phi_{N,2}|, 
\end{equation}
The noise tolerance threshold of $\mathcal{W}_c$ reads
\begin{equation}p_{\text {noise }}=\left\{\begin{array}{ll}
\left(4-4 / 2^{N / 2}\right)^{-1} & \text { even $N$}  \\
{\left[4-2\left(1 / 2^{(N+1) / 2}+1 / 2^{(N-1) / 2}\right)\right]^{-1}} & \text {odd $N$} 
\end{array}\right.\label{noise}\end{equation}
We plot $p_{\text{noise}}$ in Fig.~\ref{fig3} as a function of $N$. For large $N$, the noise tolerance is close to 0.25.

We find that our method can achieve a big improvement over the result in Eq.~(\ref{noise}), despite using the same number of measurement settings as the witness operator $\mathcal{W}_c$, which can be useful for detecting genuine  entanglement in cluster state experiments. 
We use $N$ expectation values to calculate the lower bound of the generalized geometric measure. For the cluster state mixed with a certain amount of white noise, the expectation values of the stabilizer generators are $\langle S_1\rangle=\langle S_2\rangle=...=\langle S_n\rangle=1-p_{\text{noise}}$. We find that the lower bound of GGM is positive  when $p_{\text{noise}}<0.5$ for all particle number $N$. Our result is plotted in Fig.\ref{fig3}. We see that that our method can tolerate higher noise than the witness operator $\mathcal{W}_c$.

Similar ideas can be used for the detection of entanglement in non-stabilizer states. For such states it is not possible to find $2^N$ tensor products of single-qubit operators that stabilize the state. For example, the three-qubit $W$-state $|W_3\rangle=(|001\rangle+|010\rangle+|100\rangle)/\sqrt{3}$ is a non-stabilizer state. If non-local operators are considered, $|W_3\rangle$ state is stabilized by \cite{PhysRevA.72.022340}
\begin{equation}\begin{array}{l}
S_{1}^{\left(W_{3}\right)}:=-\frac{1}{3}\left(ZZI-2 XXI-2 YIY\right) \\
S_{2}^{\left(W_{3}\right)}:=-\frac{1}{3}\left(IZZ-2 IXX-2 YYI\right) \\
S_{3}^{\left(W_{3}\right)}:=-\frac{1}{3}\left(ZIZ-2 XIX-2 IYY\right)
\end{array}\label{2727}\end{equation}
A witness operator detecting genuine three-qubit entanglement around a $|W_3\rangle$ state with only three measurement settings is constructed in Ref.~\cite{PhysRevA.72.022340}:
\begin{equation}\mathcal{W}_w=\frac{11}{3} \mathbb{I}+2 Z_{1} Z_{2} Z_{3}-\frac{1}{3} \sum_{k \neq l}\left(2 X_{k} X_{l}+2 Y_{k} Y_{l}-2 Z_{k} Z_{l}\right)\end{equation}

For the $|W_3\rangle$ state mixed with certain amount of white noise, $\rho=p_{\text{noise}} \mathbb{I}/8 +(1-p_{\text{noise}})|W_3\rangle\langle W_3| $, the witness operator $\mathcal{W}_w$ tolerates noise if $p_{\text{noise}}<4/15\approx0.27$.

In our new scheme, we choose to measure the three non-local stabilizers in Eq.~(\ref{2727}), which also requires three measurement settings $XXX$, $YYY$, and $ZZZ$. The three expectation values can be used to calculate the lower bound of generalized geometric measure. For the $|W_3\rangle$ state mixed with white noise, the measurement results of the stabilizer generators are $\langle S_1^{(W_3)}\rangle=\langle S_2^{(W_3)}\rangle=\langle S_3^{(W_3)}\rangle=1-p_{\text{noise}}$. According to our numerical calculation, the lower bound of GGM is positive if $p_{\text{noise}}<0.45$. Thus our scheme can detect genuine entanglement when $p_{\text{noise}}<0.45$, which is an improvement over the witness operator $\mathcal{W}_w$ with same number of measurement settings.

Next, we apply our new scheme of detecting entanglement to a previous cluster state experiment. In Ref.~\cite{Wunderlich_2011}, the authors prepared a four-qubit linear cluster state and measured the stabilizer generators. The expectation values are shown in Table.\ref{table2}. Using their experimental data, we can calculate the lower bound of GGM: $\varepsilon_{GGM}=0.170\pm0.004$ for the four-qubit state. Obtaining positive values for the lower bound means the state prepared in the experiment is genuine entangled.

Although previous witness operators can also detect genuine entanglement in this case, our method can tolerate higher noise. Take the four-qubit cluster state mixed with white noise for example. If $p_{\text{noise}}\approx0.4$, i.e. $\langle g_1\rangle=\langle g_2\rangle=\langle g_3\rangle=\langle g_4\rangle \approx0.6$, the witness operator $\mathcal{W}_1$ will give a positive expectation value and does not detect genuine entanglement, while our method gives a lower bound of GGM $\varepsilon_{GGM}\approx 0.1$, which means it can detect genuine entanglement in the presence of high noise.

\begin{table}\centering
\begin{tabular}{lc}
\hline Generators & Expectation values \\
\hline$g_{1}=-Z_{1} \otimes Z_{2} \otimes \mathbb{I}_{3} \otimes \mathbb{I}_{4}$ & $0.994 \pm 0.001$ \\
$g_{2}=-X_{1} \otimes X_{2} \otimes Z_{3} \otimes \mathbb{I}_{4}$ & $0.849 \pm 0.003$ \\
$g_{3}=\mathbb{I}_{1} \otimes Z_{2} \otimes X_{3} \otimes X_{4}$ & $0.937 \pm 0.003$ \\
$g_{4}=\mathbb{I}_{1} \otimes \mathbb{I}_{2} \otimes Z_{3} \otimes Z_{4}$ & $0.911 \pm 0.002$ \\
\hline
\end{tabular}
\caption{Expectation values of the stabilizer generators for the four-qubit linear cluster state in the experiment in Ref.\cite{Wunderlich_2011}. The table is reproduced from Table 1 in Ref.\cite{Wunderlich_2011}}
\label{table2}
\end{table}


In this paper we proposed a new way of witnessing entanglement, which can detect entanglement in bipartite unfaithful states. Furthermore, we extended the result to the case of multi-particle states. We showed that our scheme can tolerate higher noise than previous witness operators while using the same number of measurement settings. Our scheme is useful for detecting entanglement in experimentally prepared states. 

Although there are nonlinear witnesses which also can detect entanglement in unfaithful states~\cite{PhysRevLett.96.170502,G_hne_2007,PhysRevA.85.062327,PhysRevA.78.032326}, they usually require more measurements. Moreover, they can only be applied to bipartite systems, which means they cannot be generalized to detect genuine entanglement in multipartite states~\cite{Guhne_2009}. Moreover, our scheme can be easily adapted to different experimental settings and target states, as the type and number of measurements can be customized. In this regard, our scheme opens a new way to detect genuine entanglement based on limited information.

We thank discussions and comments from Ilan Tzitrin. We thank funding support from NSERC, CFI, ORF, Huawei Technologies Canada, MITACS, US Office of Naval Research, Royal Bank of Canada and the University of Hong Kong start-up grant.

\bibliography{apssamp}

\appendix
\section{Entanglement witness and unfaithful states}
\label{section2}

\textbf{Definition 1.} An $N$-qudit pure state $|\Psi\rangle$ is fully separable iff it can be written as
\begin{equation}
    |\Psi\rangle=\underset{i=1}{\stackrel{m}{\otimes}}|\Psi_{A_{i}}\rangle
\end{equation}
where $|\Psi_{A_i}\rangle$ is the state of the $i$-th qudit.

An $N$-qudit mixed state $\rho_s$ is fully separable iff it can be decomposed into convex mixture of fully separable pure states
\begin{equation}
\rho_s=\sum_i p_i |\Psi_i\rangle\langle\Psi_i|
\end{equation}

\textbf{Definition 2.} An $N$-qudit pure state $|\Psi\rangle$ is biseparable iff there exists a subsystem bipartition $\{A,\overline{A}\}$, where $\overline{A}$ is the complement of set $A$, such that the state can be written as
\begin{equation}
    |\Psi\rangle=|\Psi_A\rangle \otimes|\Psi_{\overline{A}}\rangle.
\end{equation}
An $N$-qudit mixed state $\rho_s$ is biseparable iff it can be decomposed into convex mixture of biseparable pure states
\begin{equation}
\rho_s=\sum_i p_i |\Psi_i\rangle\langle\Psi_i|
\end{equation}

\textbf{Definition 3.} An $N$-qudit mixed state $\rho$ is said to be \emph{genuine entangled} (or carry genuine multipartite entanglement) iff it is not a biseparable state.

\textbf{Definition 4.} A fidelity witness operator that detects genuine entanglement of a pure state $|\psi\rangle$ (and of states that are close to $|\psi\rangle$) is given by \cite{PhysRevLett.92.087902}
\begin{equation}\mathcal{W}=\alpha \mathbb{I}-|\psi\rangle\langle\psi|\label{11}.\end{equation}
The constant $\alpha$ denotes the maximum overlap between the state $|\psi\rangle$ and any biseparable state $|\phi\rangle$:
\begin{equation}\alpha=\max _{|\phi\rangle \in B}|\langle\phi \mid \psi\rangle|^{2}\end{equation}
where $B$ is the set of biseparable states. The measured value $Tr(\mathcal{W}\rho)$ is positive for any biseparable state $\rho$. On the contrary, we know for sure that $\rho$ carries genuine multipartite entanglement if we obtain a negative expectation value. The witness operator in Eq.~(\ref{11}) is called the fidelity witness because it involves measuring the fidelity between the state $\rho$ and $|\psi\rangle$. 
 
In Ref.~\cite{PhysRevLett.92.087902}, the authors implement the fidelity operator experimentally to detect entanglement in the three-qubit W state and GHZ state. In order to measure the fidelity witness in the experiment, it must be decomposed into sum of locally measurable operators~\cite{PhysRevA.66.062305}. Despite its compact mathematical form, the fidelity witness is difficult to implement in large systems because the number of measurements grows exponentially with the number of qubits~\cite{int,PhysRevLett.92.087902}, so it is only feasible in small systems.

In order to solve this problem, a large class of witness operators called \emph{fidelity-type witnesses} has been derived~\cite{PhysRevA.72.022340,Guhne_2009}. A fidelity-type witness $\mathcal{W}'$ is constructed using the following inequality:
\begin{equation}
    \mathcal{W}'-\alpha \mathcal{W}\geq0,
\end{equation}
which means $\mathcal{W}'-\alpha \mathcal{W}$ is a positive semi-definite matrix and $\alpha$ is certain positive number. Thus the witness operator $\mathcal{W}'$ can detect entanglement ($\operatorname{Tr}(\mathcal{W}'\rho)<0$) only when the fidelity witness can ($\operatorname{Tr}(\mathcal{W}\rho)<0$). Although the fidelity-type witness $\mathcal{W}'$ cannot detect as many entangled states as the fidelity witness $\mathcal{W}$, it can be implemented with fewer measurements. For example, in Ref.~\cite{PhysRevA.72.022340}, the authors constructed fidelity-type witnesses for cluster states and GHZ states, which can be implemented using two measurement settings.

Recently, researchers found a large class of states whose entanglement cannot be detected using any fidelity witnesses and therefore not with any fidelity-type witness either\cite{PhysRevLett.124.200502}. These states are called \emph{unfaithful states}. Let $\mathcal{S}$ denotes the set of separable states and $\mathcal{U}_2$  denotes the set of states that satisfy $\operatorname{Tr}(\mathcal{W}\rho)\geq0$ for any fidelity witness $\mathcal{W}$. In order to find the percentage of the set of bipartite unfaithful state $\mathcal{U}_2\backslash\mathcal{S}$ in the whole Hilbert space, they use a semidefinite programming (SDP) ansatz. The set $\mathcal{S}^1$ and $\tilde{\mathcal{U}}_2\backslash\mathcal{S}^1$ are the outer approximations of the set separable state $\mathcal{S}$ and $\mathcal{U}_2\backslash\mathcal{S}$. The percentages are calculated by sampling random bipartite states according to Hilbert-Schmidt measure~\cite{Zyczkowski_2003,PhysRevA.71.032313} and the Bures measure~\cite{Al_Osipov_2010}. The lower bound of the percentage of separable state and unfaithful state in the whole Hilbert space is shown in Table \ref{table1}. We can easily find for $d\geq3$ that almost all states in the Hilbert space are unfaithful. For $d>5$, the authors find that all states they generated are entangled but at the same time unfaithful, regardless of what metric is used to sample them. This highlights the importance of characterizing
the entanglement of unfaithful qudit states beyond relying on fidelity witnesses.

\begin{table}

\centering
\begin{tabular}{lccccc}
\hline \hline $d$ & $\mathcal{S}^{1}(H S)$ & $\tilde{\mathcal{U}}_{2} \backslash \mathcal{S}^{1}(H S)$ & $\mathcal{S}^{1}(B)$ & $\tilde{\mathcal{U}}_{2} \backslash \mathcal{S}^{1}(B)$ \\
\hline  2 & 24.2 \% & 21.2 \% & 7.4 \% & 15.4 \% \\
3 & 0.01 \% & 94.5 \% & 0 \% & 54.8 \% \\
4 & 0 \% & 100 \% & 0 \% & 97 \% \\
5 & 0 \% & 100 \% & 0 \% & 100 \% \\
\hline \hline 
\end{tabular}
\caption{The lower bound on the percentage of bipartite separable states( $\mathcal{S}^1$) and unfaithful states ($\tilde{\mathcal{U}}_2\backslash\mathcal{S}^1$) in the whole Hilbert space in small dimensions. This table is reproduced from TABLE. I in Ref.\cite{PhysRevLett.124.200502}. The percentages are calculated in the Hilbert-Schmidt(HS) metric and Bures(B) metric. For $d>2$, most states in Hilbert space are both entangled and unfaithful.}
\label{table1}
\end{table}

The authors also find that most pure entangled state remained entangled but become unfaithful when a certain amount of white noise is added. Consider the Bell state $|\Psi_2\rangle=(|00\rangle+|11\rangle)/\sqrt{2}$ embedded in $d \times d$-dimensional Hilbert space:
\begin{equation}
    \rho=p\frac{\mathbb{I}}{d^2}+(1-p)|\Psi_2\rangle\langle\Psi_2|
    \label{33}
\end{equation}
\begin{figure}
    \centering
    \includegraphics[width=0.5\textwidth]{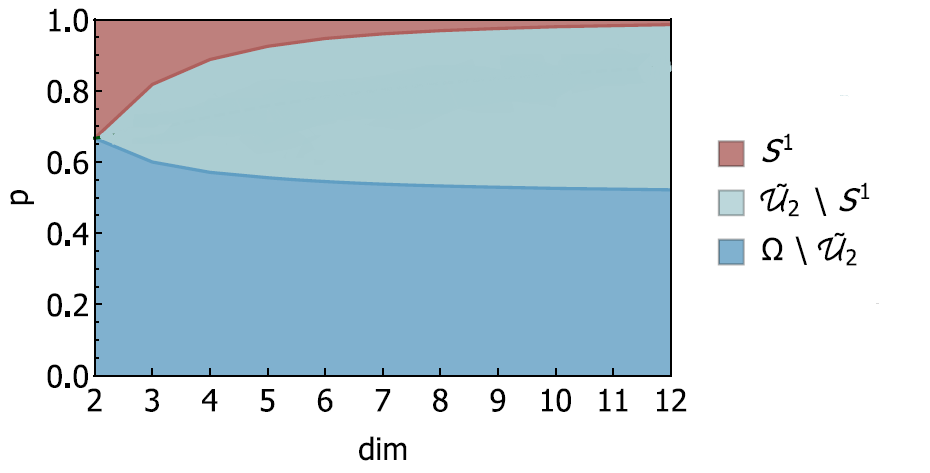}
    \caption{The unfaithful and entangled region of the family of states $\rho$ from Eq.~(\ref{33}). This figure is reproduced from FIG. 2 in Ref.\cite{PhysRevLett.124.200502}. The red region corresponds to the set of separable states. The green region indicates the set of unfaithful and entangled states. The blue region indicates the set of faithful states, whose entanglement can be detected using fidelity witness. }
    \label{noi}
\end{figure}
where $p$ is the amount of white noise. It is easy to prepare such a state by drawing a random state from either a completely mixed state or the state $| \Psi_2 \rangle$ with a probability $p$ and $1-p$, respectively. For $d>2$, there exists a large parameter regime where the state $\rho$ is entangled and unfaithful. This phenomenon is shown in Fig. \ref{noi}. What's more, they find numerically that all sampled pure entangled states are unfaithful and entangled for a certain range of white noise. The only exception they find is the maximally entangled state $1 / \sqrt{d} \sum_{i=0}^{d-1}|i i\rangle$ subjected to white noise.

\section{Quantifying entanglement with simple measurements}
\label{section3}
In Ref.~\cite{PhysRevLett.98.110502,Eisert_2007}, the authors present a method to estimate entanglement with the measured expectation value of witness operators. In this section we briefly summarize their results. 

\begin{figure}
    \centering
    \includegraphics[width=0.5\textwidth]{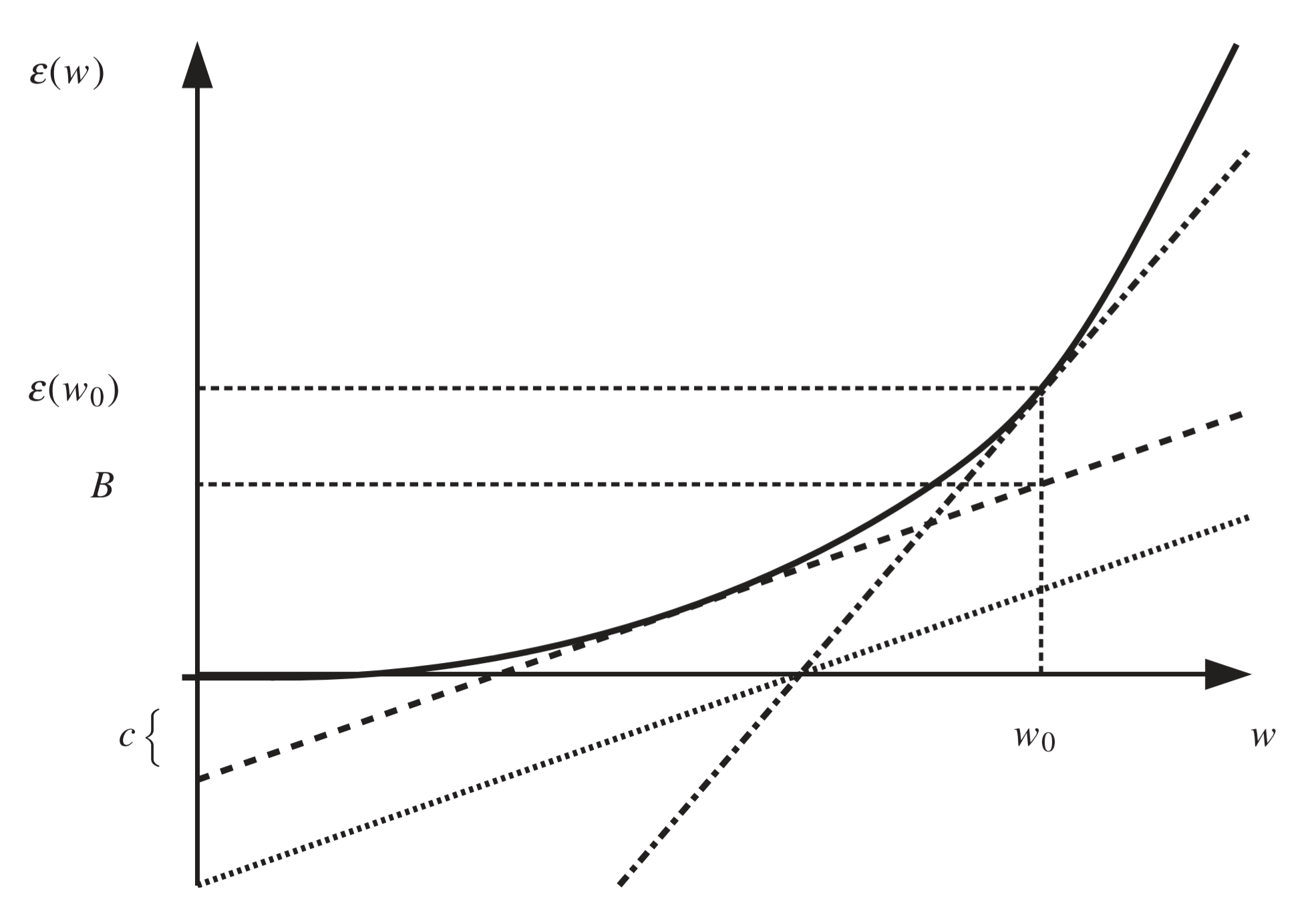}
    \caption{A schematic view of the method of estimating entanglement lower bound. This figure is reproduced from Ref.\cite{PhysRevLett.98.110502}. The lower bound $\varepsilon(w)$ is a convex function. We assume that $w_0$ is the measured expectation value. The dotted line corresponds to a general estimation. First we perform the minimization on the intercept $c$ and the linear bound becomes the dashed line. We can get a better estimation of the entanglement lower bound, which is $B$. By varying the slope $r$ we arrive at the dash-dotted line, which gives the best estimation of the lower bound $\varepsilon(w_0)$.}
    \label{fig1}
\end{figure}

Suppose we measure $n$ witness operators $\{\mathcal{W}_1,\mathcal{W}_2,...\mathcal{W}_n\}$ on the state in the experiment and their expectation values are $w_1,w_2,...w_n$. We hope to find the lower bound of entanglement based on our measurement results:
\begin{equation}\varepsilon\left(w_{1}, \ldots, w_{n}\right)=\inf _{\rho}\left\{E(\rho) \mid \operatorname{Tr}\left(\rho \mathcal{W}_{k}\right)=w_{k}\right\}\label{1010},\end{equation}
where the infimum is understood as the infimum over all states compatible with the measurement data. We note that $\varepsilon\left(w_{1}, \ldots, w_{n}\right)$ is a convex function; the proof of this is simple and can be found in the appendix.
We consider linear bounds of the type
\begin{equation}\varepsilon(w) \geq  r \cdot w-c, \label{66}\end{equation}
where $r$ and $w$ are both $n$-dimensional vectors and the dot denotes the dot product. For the one-dimensional case, the two bounds are plotted in Figure~\ref{fig1}. To find the best estimation of the entanglement lower bound $\varepsilon(w)$, the intercept $c$ should be as small as possible. By the definition of $\varepsilon(w)$, we note that Eq.~(\ref{66}) can be expressed as $E(\rho)\geq r \cdot w-c$, so we arrive at the inequality for $c$.
\begin{equation}c \geq \sum_{k} r_{k} \operatorname{Tr}\left(\rho \mathcal{W}_{k}\right)-E(\rho)\end{equation}
This indicates that we should perform maximization over the whole state space and choose $c$ to be the supremum of the right-hand side (RHS), which only depends on the operator $\mathcal{W}=\sum_k r_k \mathcal{W}_k$. Thus the supremum of the RHS can be expressed as the function $\hat{E}(\mathcal{W})$, where
\begin{equation}\hat{E}(\mathcal{W})=\sup _{\rho}\{\operatorname{Tr}(\rho \mathcal{W})-E(\rho)\}\label{1515}.\end{equation}
After obtaining the minimum value of $c$, we vary the slope $r$ to arrive at the best possible bound $\varepsilon(w)$, as we have shown in Fig.~\ref{fig1}:
\begin{equation}\varepsilon(w)=\sup _{r}\left\{\sum_k r_k w_k-\hat{E}\left(\sum_{k} r_{k} \mathcal{W}_{k}\right)\right\}\label{99}\end{equation}
Note that Eq.~(\ref{99}) shows that $\varepsilon(w)$ is the Legendre transformation of the function $\hat{\varepsilon}(r)=\hat{E}(\sum_k r_k\mathcal{W}_k)$.

The entanglement measure  $E(\rho)$ is often defined through the convex roof construction~\cite{plenio2005introduction}
\begin{equation}E(\rho)=\min_{p_i,|\psi_i\rangle} \left\{\sum_{i} p_{i} f\left(|\psi_{i}\rangle\right): \sum_{i} p_{i} |\psi_i\rangle\langle\psi_i|=\rho\right\}\label{1313},\end{equation}
where $f$ can be chosen as the geometric measure of entanglement, entanglement of formation, etc.. In this way the function $\hat{E}(\mathcal{W})$ can be evaluated by:
\begin{equation}\begin{aligned}
\hat{E}(\mathcal{W})=& \sup _{\rho}\left\{\operatorname{Tr}(\rho \mathcal{W})-\inf _{p_{i},\left|\psi_{i}\right\rangle} \sum_{i} p_{i} E\left(\left|\psi_{i}\right\rangle\right)\right\} \\
=&\sup_{p_{i}} \sup _{\left|\psi_{i}\right\rangle}\left\{\sum_{i} p_{i}\left\{\left\langle\psi_{i}|\mathcal{W}| \psi_{i}\right\rangle-E\left(\left|\psi_{i}\right\rangle\right)\right\}\right\} \\
=& \sup _{|\psi\rangle}\{\langle\psi|\mathcal{W}| \psi\rangle-E(|\psi\rangle)\}.
\end{aligned}\label{1414}\end{equation}
Thus the process of calculating the lower bound can be summarized as follows: First  we choose what kind of entanglement measure to use and calculate minimum value of the intercept $c$. Then, we do the optimization over the slope $r$ according to Eq.~(\ref{99}) to get the lower bound of entanglement $E(\rho)$.

\section{calculating the lower bound of geometric measure and generalized geometric measure}
\label{section4}
In this section we generalize the scheme in the previous section to calculate the lower bound of entanglement with measurement results of any Hermitian operators. Moreover, we show the numerical calculation of lower bound of entanglement based on a set of measurement results.

Suppose we would like to measure a set of Hermitian operators $A_1,A_2,\dots,A_n$ in an experiment and use the measured values $a_1,a_2,\dots,a_n$ to calculate the lower bound of entanglement of the state $\rho$:
\begin{equation}\varepsilon\left(a_{1}, \ldots, a_{n}\right)=\inf _{\rho}\left\{E(\rho) \mid \operatorname{Tr}\left(\rho {A}_{k}\right)=a_{k}\right\}\end{equation}
Simply replacing the witness operators $\{\mathcal{W}_1,\mathcal{W}_2,...\mathcal{W}_n\}$ in Eq.(\ref{1010}) with Hermitian operators $\{{A}_1,{A}_2,...A_n\}$, the lower bound $\varepsilon(a_1,a_2,...a_n)$ can be calculated using the scheme summarized in Section III. In this case, the maximum intercept in Eq.~(\ref{1515}) can be rewritten as
\begin{equation}\hat{E}(\mathcal{A})=\sup _{\rho}\{\operatorname{Tr}(\rho \mathcal{A})-E(\rho)\},\end{equation}
where $\mathcal{A}=\sum_k r_k A_k$. If the entanglement measure is defined through the convex roof construction, $\hat{E}(\mathcal{A})$ can be evaluated using
\begin{equation}
    \hat{E}(\mathcal{A})= \sup _{|\psi\rangle}\{\langle\psi|\mathcal{A}| \psi\rangle-E(|\psi\rangle)\}
    \label{1919}.
\end{equation}
Finally, the lower bound $\varepsilon(a_1,a_2,...,a_n)$ can be calculated using
\begin{equation}\varepsilon(a_1,a_2,...,a_n)=\sup _{r}\left\{\sum_k r_k a_k-\hat{E}\left(\sum_{k} r_{k} A_{k}\right)\right\}\label{2020}\end{equation}
which is a modified version of Eq.~(\ref{99})

In order to calculate the entanglement lower bound, we should choose the entanglement measure $E(\rho)$ first. For extending to the multipartite case, the entanglement measure must be able to quantify multipartite genuine entanglement. There exist many multipartite entanglement measures, including the Schmidt measure, relative entropy of entanglement, and the geometric measure~\cite{DBLP:journals/qic/PlenioV07}. Although the Schmidt measure is widely used in theoretical analysis of the entanglement of cluster states, it is not robust to noise, i.e., it may vary a lot if there is any small deviation in the quantum state. Since the relative entropy is not constructed through the convex roof, $\hat{E}(\mathcal{A})$ is harder to evaluate.


In Ref.~\cite{PhysRevLett.98.110502}, the authors show the calculation of the lower bound of the Entanglement of Formation and the geometric measure. In this paper, we use the geometric measure as a measure of bipartite entanglement. The geometric measure is an entanglement monotone~\cite{PhysRevA.68.042307} and is relatively easy to calculate~\cite{Teng_2017,PhysRevA.84.022323}. For pure states, the geometric measure is defined as one minus the maximal squared overlap with separable states.
\begin{equation}E_{G}(|\psi\rangle)=1-\sup _{|\phi\rangle=|a\rangle|b\rangle|c\rangle\cdots}|\langle\phi \mid \psi\rangle|^{2}\end{equation}
For mixed states, the geometric measure of entanglement is defined via the convex roof construction in Eq.~(\ref{1313}). By replacing the $E(|\psi\rangle)$ term in Eq.(\ref{1919}) with the geometric measure, we get $\hat{E}_{G}(\mathcal{A})$ for bipartite states
\begin{equation}\hat{E}_{G}(\mathcal{A})=\sup_{|\psi\rangle}\sup_{|\phi\rangle=|a\rangle|b\rangle}\{\langle\psi|(\mathcal{A}+|\phi\rangle\langle\phi|)| \psi\rangle-1\}\label{1818}\end{equation}

The function $\hat{E}_{G}(\mathcal{A})$ can be evaluated by doing optimization over all biseparable states $|\phi\rangle=|a\rangle|b\rangle$. After that, the entanglement lower bound $\varepsilon(a_1,a_2,...,a_n)$ can be calculated by optimizing the slope $r$ using Eq.~(\ref{2020}). The numerical algorithm for the optimization is presented in the appendix. By calculating the lower bound of entanglement, we can determine whether the bipartite state is entangled or not.

However, things become different when detecting genuine entanglement in multipartite states. The geometric measure of a multipartite state is positive if the state is not fully separable, so obtaining a positive lower bound does not means the state carries genuine entanglement. Here we use another kind of entanglement measure, the generalized geometric measure (GGM)~\cite{PhysRevA.94.022336,PhysRevA.81.012308}. The GGM of a pure state $|\psi\rangle$ is defined by
\begin{equation}
E_{GGM}(|\psi\rangle)=1-\sup _{|\phi\rangle=|\phi_A\rangle|\phi_{\overline{A}}\rangle}  |\langle\phi \mid \psi\rangle|^2
\end{equation}
where the optimization is done over all biseparable states $|\phi\rangle$. An equivalent form of the equation is
\begin{equation}E_{GGM}\left(\left|\psi\right\rangle\right)=1-\max \lambda_{A: \overline{A}}^{2}\end{equation}
where $\lambda_{A: \overline{A}}$ is the maximal Schmidt coefficient for any bipartite split $A:\overline{A}$ of the state $|\psi\rangle$. For mixed states, $E_{GGM}$ is constructed using the convex-roof construction. $E_{GGM}$ is positive iff the state carries genuine entanglement. The calculation of lower bound of $E_{GGM}$ is similar to that of $E_G$, as we only have to change the optimization region from fully separable states to biseparable states. If we get a positive value for the lower bound of $E_{GGM}$, we know the state is genuine entangled.
\section{cluster states}
A cluster state is a type of highly entangled state of multiple qubits. It was first introduced in \cite{PhysRevLett.86.910}. One important application of cluster states is as resource states in measurement-based computation, which was proposed in Ref.~\cite{PhysRevLett.86.5188}. Cluster states have been carefully studied in the past two decades. Experimentalists have also tried to realize cluster states for the ultimate goal of measurement-based quantum computation~\cite{PhysRevLett.99.120503,PhysRevA.94.032327,nphys,PhysRevLett.95.210502,Schwartz434}.

\begin{figure}
    \centering
    \includegraphics[width=0.3\textwidth]{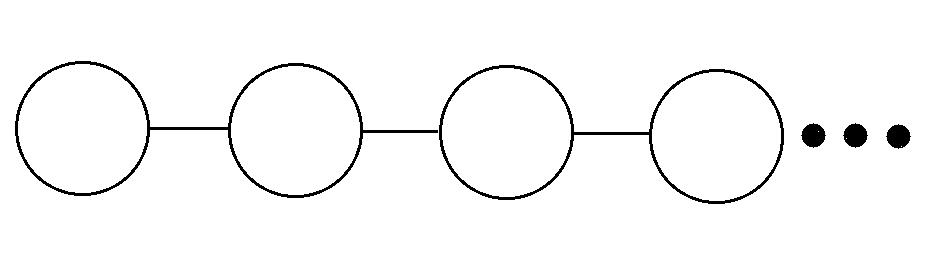}
    \caption{An $N$-partite linear cluster state.}
    \label{cluster}
\end{figure}

An $N$-partite, $d$-level linear cluster state $|\phi_{N,d}\rangle$ is shown in Fig.~\ref{cluster}. It is uniquely defined by the eigenvalue equations
\begin{equation}X_{a} \bigotimes_{b \in \mathcal{N}(a)} Z_{b}\left|\phi_{N, d}\right\rangle=\left|\phi_{N, d}\right\rangle\end{equation}
where
\begin{equation}\mathcal{N}(a)=\left\{\begin{array}{l}
\{2\}, \quad a=1 \\
\{N-1\}, \quad a=N \\
\{a-1, a+1\}, \quad a \notin\{1, N\}
\end{array}\right.\end{equation}
The $X$ and $Z$ are generalized Pauli matrices, which are introduced in Ref.~\cite{PhysRevA.68.062303}. In the stabilizer formalism, the $N$ stabilizer generators are $\mathcal{S}_k=X_a \bigotimes Z_b$, where $k=1, \dots ,N$.

\section{Numerical optimization algorithms}
A simple optimization scheme for evaluating $\hat{E}_G(\mathcal{W})$ in Eq.~(\ref{1818}) is introduced in Ref.~\cite{PhysRevLett.98.110502}. Here we briefly summarize their scheme. If $|\phi\rangle$ is fixed, we can perform optimization by choosing $|\phi\rangle$ as the eigenvector corresponding to the largest eigenvalue of $(\mathcal{A}+|\phi\rangle\langle\phi|)$. If $|\psi\rangle$ is fixed, the method of finding the closest separable state is as follows: First we fix $|b\rangle,|c\rangle...$ and perform optimization on $|a\rangle$; then, we fix $|a\rangle,|c\rangle...$ and perform optimization on $|b\rangle$. After several iterations we can arrive at the optimal separable state $|\phi\rangle$. Finally, the optimization on $|\phi\rangle$ and $|\psi\rangle$ can be iterated, thus converging to the value of $\hat{E}_G(\mathcal{A})$. 

However, in numerical calculations we find that the scheme above for finding the optimal $|\phi\rangle$ often fails. In many cases the iterations do not converge. The problem of numerically calculating the geometric measure of entanglement (finding the optimal  $|\phi\rangle$) was carefully studied before~\cite{Teng_2017,PhysRevA.84.022323}. Here we use the scheme introduced in \cite{Teng_2017}.
  
The Tucker decomposition~\cite{wikiTuck} of tensor $T_{m n p \ldots z}$ is defined as
\begin{equation}T_{m n p \ldots z}=\sum \lambda_{\alpha \beta \gamma \ldots \omega} a_{\alpha m} \circ a_{\beta n} \circ a_{\gamma p} \cdots \circ a_{\omega z}\end{equation}
where $\lambda$ and $a$ are tensors. The objective function of a rank-$k$ approximation of the tensor $T_{m n p \ldots z}$ can be written as
\begin{equation}d=\min \left\|T_{m n \cdots p}-\sum_{i=1}^{k} \lambda_{i} a_{i m} \circ a_{i n} \cdots \circ a_{i p}\right\|\end{equation}

The problem of finding the separable state $|\phi\rangle$ with maximum overlap with $|\psi\rangle$ is the same as finding the rank-1 approximation of the tensor $T$. There are numerous algorithms that can be used for the  rank-$k$ approximation. The Alternate Least Squares algorithm is one of the most popular approaches. We will not discuss the details of the algorithms here; a complete survey of the algorithms can be found in Ref.~\cite{doi:10.1137/07070111X}. There are also numerous existing code packages that can be utilized on different coding platforms, such as C++ and MATLAB. For this paper, we use the MATLAB Tensor Toolbox 2.6 developed by Sandia National Laboratories~\cite{TTB_Software}. This package is already developed and available online.
  
We can perform the optimization in Eq.~(\ref{1313}) in the following way: By using the \texttt{eigs} function in MATLAB, we can compute the eigenvector $|\psi\rangle$ of $(\mathcal{W}+|\phi\rangle\langle\phi|)$ corresponding to the maximum eigenvalue. Then, we fix $|\psi\rangle$ and compute the separable state wavefunction $|\phi\rangle$ with the maximal overlap using the method introduced above. After that, we iterate the two steps. The convergence of such iteration is very fast. In most cases, the numerical error can be reduced to less than $10^{-4}$ in less than 4 iterations.

After calculating $\hat{E}_G(\mathcal{W})$, the minimum value of $c$, we optimize the slope $\vec{r}$. As $\varepsilon(w)$ is a convex function of $w$, the Legendre transformation $\hat{\varepsilon}(\vec{r})$ is also a convex function of $\vec{r}$~\cite{Convex}. Therefore, the Gradient Descent algorithm can be applied to perform optimization. For each iteration the slope $\vec{r}$ is updated as
\begin{equation}
    \vec{r}\rightarrow\vec{r}+\eta\nabla\hat{\varepsilon}(\vec{r}),
\end{equation}
where $\eta$ is the learning rate. After several iterations, $\vec{r}$ will reach the global maximum point of $\hat{\varepsilon}(\vec{r})$, which is the best lower bound of entanglement.
\nocite{*}


\end{document}